# Xeno Amino Acids: A look into biochemistry as we don't know it

Sean M. Brown[*,1,†], Christopher Mayer-Bacon[1,2], & Stephen Freeland[1,†]

[1]University of Maryland, Baltimore County: Department of Biological Sciences, Baltimore, Maryland, United States
[2]United States Food and Drug Administration, Silver Spring, Maryland, United States
*Author to whom correspondence should be addressed.
[†]These authors contributed equally to this work.

**Keywords:** Astrobiology; Amino Acid; Review; Xenobiology; Peptide Biochemistry; Abiogenesis; Evolution

## Abstract

Would another origin of life resemble Earth's biochemical use of amino acids? Here we review current knowledge at three levels: 1) Could other classes of chemical structure serve as building blocks for biopolymer structure and catalysis? Amino acids now seem both readily available to, and a plausible chemical attractor for, life as we don't know it. Amino acids thus remain important and tractable targets for astrobiological research. 2) If amino acids are used, would we expect the same L-alpha- structural subclass used by life? Despite numerous ideas, it is not clear why life favors L-enantiomers. It seems clearer, however, why life on Earth uses the shortest possible (alpha-) amino acid backbone, and why each carries only one side chain. However, assertions that other backbones are physicochemically impossible have relaxed into arguments that they are disadvantageous. 3) Would we expect a similar set of side chains to those within the genetic code? Not only do many plausible alternatives exist and evidence exists for both evolutionary advantage and physicochemical constraint for those encoded by life. Overall, as focus shifts from amino acids as a chemical class to specific side chains used by post-LUCA biology, the probable role of physicochemical constraint diminishes relative to that of biological evolution. Exciting opportunities now present themselves for laboratory work and computing to explore how changing the amino acid alphabet alters the universe of protein folds. Near-term milestones include: a) expanding evidence about amino acids as attractors within chemical evolution; b) extending characterization of other backbones relative to biological proteins; c) merging computing and laboratory explorations of structures and functions unlocked by xeno peptides.

# Introduction

A key question for astrobiology is whether life originating elsewhere in the universe would share similar biochemistry to that of life on Earth? Here, we narrow the challenging focus of that question to the topic of amino acids.

A foundational step of early biological evolution was to establish a genetically encoded 'alphabet' comprising 20 different amino acids. Since then, the greatest deviations in ~3.5 billion years have been the addition of a 21st amino acid (Selenocysteine, Sec) within some lineages of bacteria (Böck et al., 1991), archaea (Rother et al., 2010) and eukaryotes (Kivenson et al., 2021); and a 22nd, pyrrolysine (Pyl), in two of these three domains (archaea: Sun et al., 2021; and bacteria: Brugère et al., 2018). This clear process of evolutionary extension from 20 to 22 (Ambrogelly et al., 2007) complements evidence that the canonical set of 20 is itself an outcome of biological evolution rather than a chemical prerequisite for life to begin. And yet it is the canonical alphabet of 20, a foundation of biological, biochemical, and biomedical research, where knowledge has accumulated. The very architecture built to facilitate contemporary biological research constrains how little we know about possibilities for biochemistry beyond this set. It is still new biotechnology to develop laboratory protocols for manipulating and analyzing biological proteins beyond the genetically encoded 20 amino acids (e.g. compare Young & Schultz, 2010 with Opuu & Simonson, 2023). The data and tools of bioinformatics are mostly built around the assumption that any site within a biological protein can exist in one of 20 states. However, such a fundamental feature of life on Earth offers tempting potential for developing tractable, focused ideas about agnostic biosignatures. Whatever can be established about the likelihood of life elsewhere in the universe using amino acids or, better yet, about the characteristics of a "life-sustaining set," is a direct and significant contribution to current astrobiology.

Here, we review amino acids by revisiting and expanding three questions first introduced by Weber & Miller (1981): (1) Why does life on Earth use amino acids, rather than some other class of molecule? (2) Why does it use L-$\alpha$-amino acids rather than other structural sub-classes? and (3) Why does the post-LUCA genetic code comprise 20-22 specific side chains? But whereas Weber and Miller approached the topic as chemists - summarizing what some call "bottom-up" thinking (Preiner et al., 2020) - we approach the topic from biology, reasoning "top-down" as we work backward from life as we know it. In addressing each question, we focus on what is known and what is unknown about "xeno" amino acids - those from beyond the genetically encoded alphabet of 20.

## BOX 1. THE ROLE OF AMINO ACIDS IN TERRESTRIAL BIOLOGY

Everything alive today constructs metabolism primarily as a network of genetically encoded proteins. Each protein is a polymerized sequence of amino acids. In 1972, Christian Anfinsen was awarded the Nobel Prize in Chemistry for demonstrating that a protein's primary sequence (i.e. which members of the amino acid alphabet are joined together and in what order) determines how a linear polymer folds into a three dimensional conformation (Anfinsen, 1973). Since LUCA, all life on Earth genetically encodes 20 of these amino acids, although some lineages are evolving to add Selenocysteine and/or Pyrrolysine. Each of these 20 (+2) genetically encoded amino acids (**A**) is defined by a constant backbone (**B**): an amine (-NH2) at one end, a carboxyl (-COOH) at the other, and an "alpha" carbon atom between these two functional groups. This carbon atom carries a variable side chain (R-group), and differences between these side-chains distinguish each amino acid. Proteins are formed when a covalent, peptide bond links the carboxyl group (-COOH) of one amino acid to the amino group (-NH2) of another (**C**). The resulting thread-like backbone of every protein contains a series of rotatable bonds ($\phi$ and $\Psi$) within this peptide chain (**D**). The Phi ($\phi$) and Psi ($\Psi$) angles around each alpha carbon define any given protein's 3-dimensional structure (**E**).

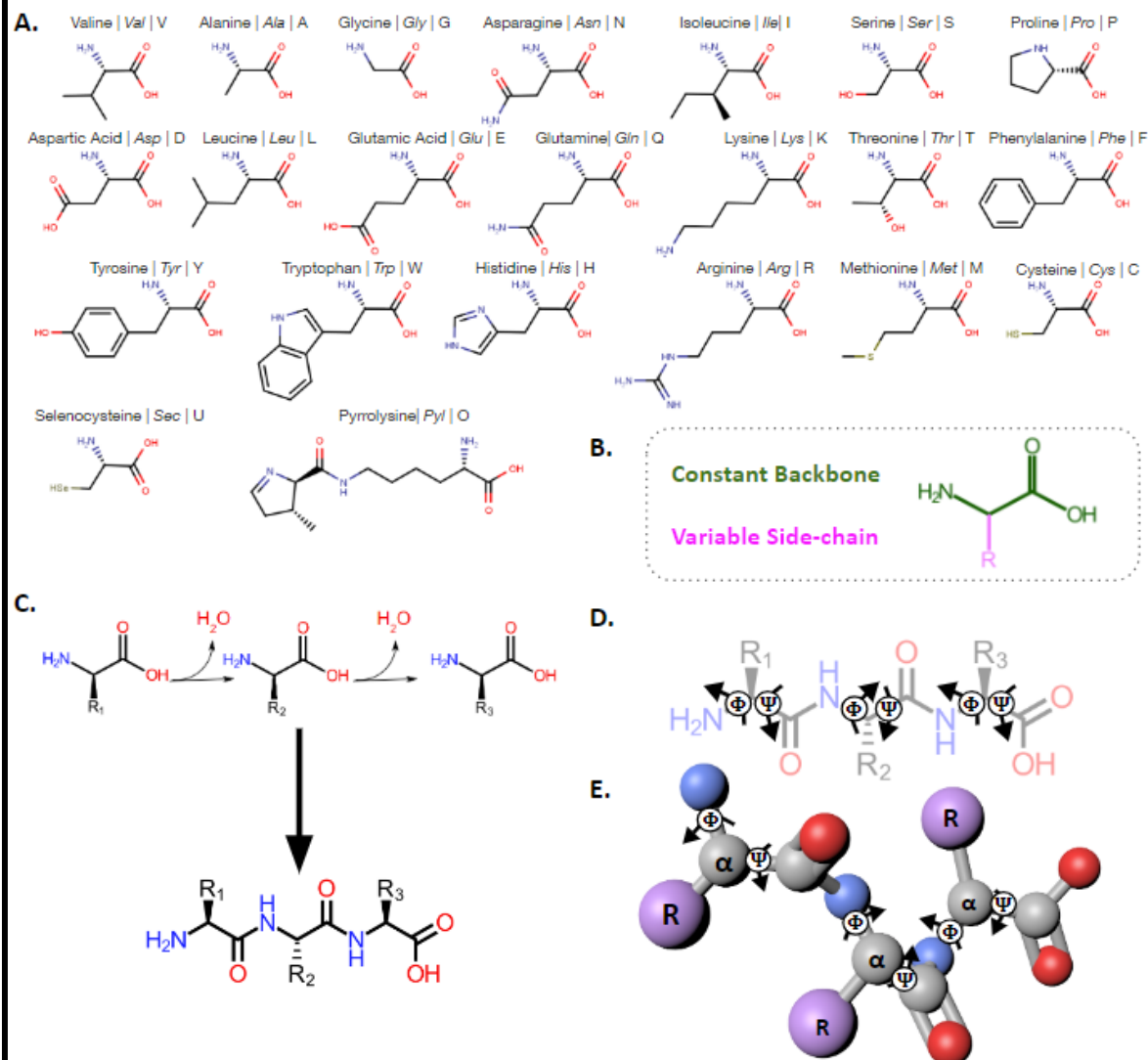

# Would a xeno biochemistry use **amino acids**?

An exploration of amino acids' relevance to xeno biochemistry can usefully begin with one, simple observation: the standard or canonical alphabet of 20 genetically encoded amino acids is extremely good at what it does. Protein-based metabolism constructed with this one set of molecular building blocks has diversified so successfully that contemporary scientists remain actively engaged in finding environmental limits to life on Earth (e.g. Merino et al., 2019: Table 4). Genetically encoded proteins sustain biology, for example, in polar volcanoes (Bendia et al., 2018), nuclear-contaminated sites (Fredrickson et al., 2004), and "toxic" acid-mine drainage (Baker & Banfield, 2003). A current understanding is that "*water activity appears to be the single key parameter controlling the biospace of Earth's life, and numerous other parameters limiting life (e.g., temperature and salinity) are, in fact, acting on the availability of water*" (Merino et al., 2019). Indeed, life on Earth is now recognized to flourish across a range of conditions that overlap significantly with extraterrestrial environments (*ibid*: Fig. 2). The standard amino acid alphabet has even proven sufficient to sustain life for three years on the exterior of the International Space Station (Kawaguchi et al., 2020) and evidence is growing that Earth life could travel between different planetary bodies (Danko et al., 2021)!

Of course, amino acids' impressive potential for constructing polymer catalysts is merely consistent with, not evidence for, their likely use within a xeno biochemistry. An important, complementary question follows: would an independent origin of life "discover" this type of organic molecule? In the mid-20th century, "spark tube" experiments produced the first suggestion of an affirmative answer by simulating physicochemical conditions thought to represent a prebiotic planet Earth (Miller 1953; Miller & Urey, 1959). A circulating mixture of water, methane, hydrogen, and ammonia provided with energy in the form of heat and an electric spark was shown to produce amino acids, among many other organic compounds. This direct connection between the abiotic universe and fundamental biochemistry inspired an entire literature which, in short, reveals that variations in both energy source and reactants change only the quantity of amino acids produced, and the diversity of side chains, not their presence/absence as a class of chemical structure (see Cleaves, 2010 for a thorough review, but with plentiful, continuing research e.g. Kebukawa et al., 2017; Pietrucci et al., 2018; Magrino et al., 2021).

Since the 1970's, this direct connection between the abiotic universe and biochemistry has been affirmed by analysis of carbonaceous meteorites (Kvenvolden, 1971), the organic chemistry of which provides a natural analog to laboratory simulations (Pizzarello and Shock, 2010; Burton et al., 2012; Glavin et al., 2018). Indeed, advances in instrumentation reveal an increasingly diverse repertoire of amino acids among the organic compounds found both within newly discovered meteorites (e.g. Elsila et al., 2016; Simkus et al., 2019) and reanalysis of those studied previously (e.g. Koga and Naraoka, 2017). Recent advances in space sciences are now removing the need to wait for meteorites to fall to Earth as earlier this year, fifteen different amino acids were identified *in situ* on the Ryugu asteroid (Naraoka et al., 2023) formed by multiple reaction pathways (Parker et al., 2023). Echoing the broad findings of prebiotic simulations, it is not the presence/absence of amino acids that changes in these meteorites, but rather "*The abundances [that] vary significantly [according to] different degrees of secondary alteration processes including thermal and aqueous alteration*" (Aponte et al., 2020). Contemporary science is increasingly clear that abiotic organic chemistry synthesizes amino acids almost anywhere that sufficient energy melts ice into water in the presence of organic carbon and nitrogen.

This cosmic ubiquity distinguishes amino acids from other fundamental components of biochemistry (Fig. 1). In addition to proteins, life as we know it comprises genetic material in the form of polymerized nucleotide sequences and is encapsulated within lipid membranes. Neither lipids nor nucleotides form easily under prebiotic conditions. Certainly, fatty acids occur and could potentially play a role as forerunners to lipids (Deamer 2017; Todd et al., 2022), and nucleobases, a subcomponent of nucleotides, also occur (Oró, 1961; Okamura et al., 2019) but prebiotic synthesis of nucleotides themselves is far more controversial (Orgel, 2004; Engelhart and Hud, 2010; Fine & Pearlman, 2023). Broadly speaking, the difference can be understood from the number and types of atoms involved: amino acids in general, and those produced by abiotic synthesis in particular, comprise fewer atoms than lipids or nucleotides (Table 1). Not only do larger molecules imply the need for more atoms to find and react with one another in the absence of any guiding enzyme, but the addition of each new heavy atom brings exponentially expanding structural combinations (Meringer, 2013). Thus ribose ($C_5H_{10}O_5$) is formed by one of the oldest organic syntheses known to science (Boutlerow, 1861), but in the absence of catalysis (Ricardo et al., 2004), total synthesis yield divides between countless other structures that share a similar chemical formula (Decker, 1982; Shapiro, 1988). Such simple generalizations of course ignore many sophisticated considerations, most notably reaction pathway dynamics and a role for non-biological catalysts, but their usefulness is supported by noting that nucleobases and fatty acids fall within a molecular weight range similar to that of abiotically plausible amino acids, while nucleotides and lipids do not (Table 1).

      Beyond mere atom counts, amino acids distinguish themselves by chemical composition from the other fundamental components of biochemistry. All components comprise just six chemical elements (C, H, N, O, P, and S). Excluding only the noble gasses (He, Ne, and Ar), four of these six 'biochemical' elements (C, H, O, and N) are the most abundant atoms in the universe and are sufficient to produce 18 of the 20 genetically encoded amino acids. The remaining two amino acids require only the addition of sulfur, which follows close behind in terms of abundance (Asplund et al., 2006). In contrast, both nucleotides and biological membrane lipids incorporate phosphorus, which is generally far less abundant than C, H, O, N, or S. Again, this argument ignores a host of more sophisticated considerations, such as microenvironments that may have delivered phosphorus to an origin of life (Pasek, 2008). Overall it is clear, however, that prebiotic synthesis of RNA remains persistently more challenging than that of amino acids. The world's leading research here continues to search for whatever it is that all previous efforts have missed. Either the right kind of mineral surface was needed to catalyze the pathways which form and derivatize ribose (Jerome et al., 2022), or no minerals are required because nucleobases lacked any backbone in life's earliest stages (Schuster et al., 2021) or some combination of physicochemical conditions, not yet tried, is the missing answer. Answers here are potentially endless: recent examples include photochemistry (Green et al., 2021) or the sort of cyclical self-purification of RNA from within a more heterogeneous polymer (Kim et al., 2021) that we describe below for the case of amino acids versus hydroxy acids (Forsythe et al., 2015). In this context, it is noteworthy that at least some of the ingenious chemistry developing here is overtly motivated by the perception that an RNA world paradigm provides "*a mandate for chemistry to explain how RNA might have been generated prebiotically on the early Earth*" (Anastasi et al., 2007), and must thus be balanced against serious arguments that RNA might instead be a product of early biological evolution rather than a prerequisite (Engelhart & Hud, 2010; Freeland 2022; Fine & Pearlman, 2023).

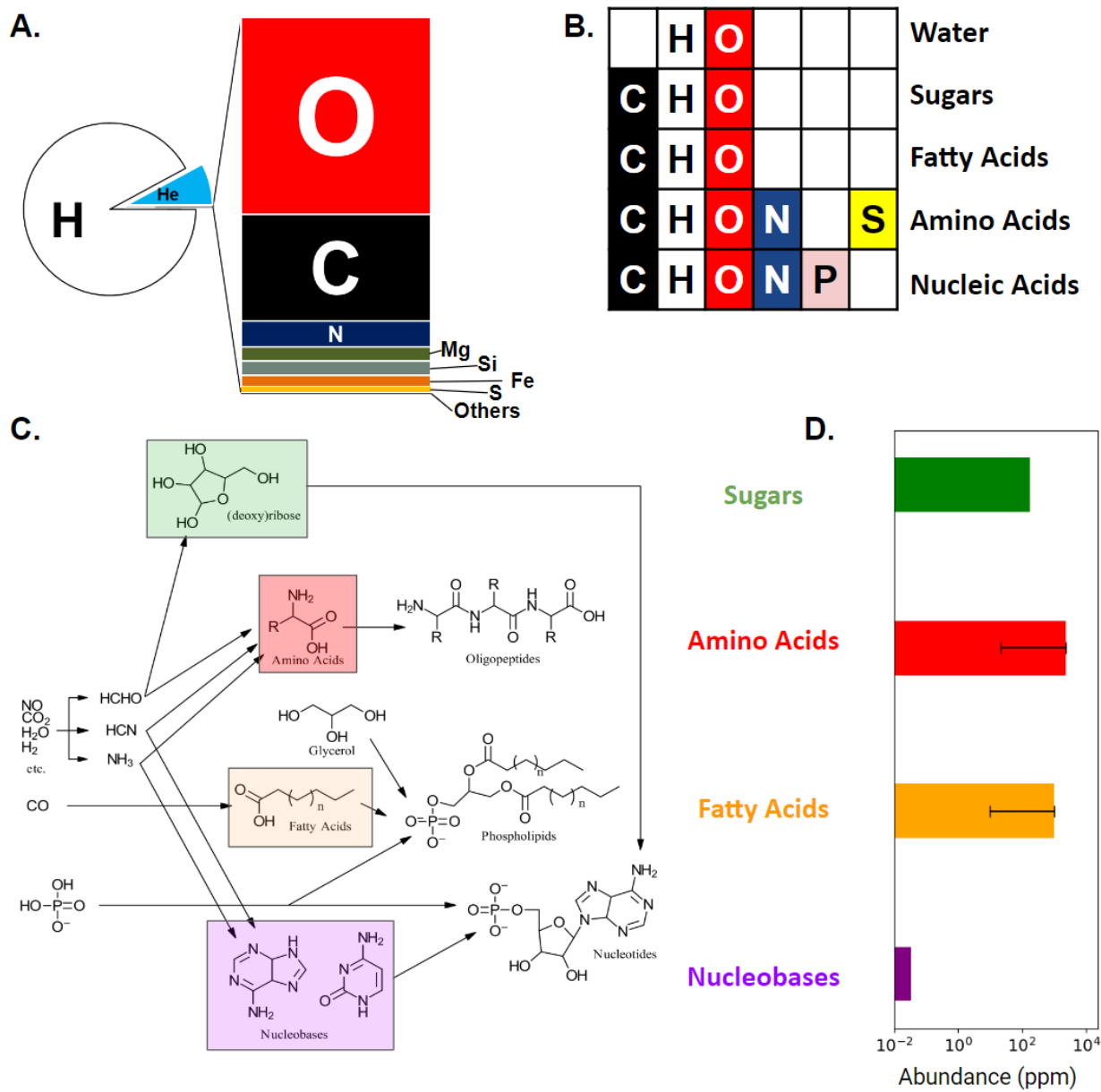

**Figure 1**. Life's fundamental biochemistry comprises just six chemical elements (carbon, nitrogen, hydrogen, oxygen, sulfur, and phosphorus). **(A)** The atomic composition of the Milky Way Galaxy (Croswell, 1996) is primarily dominated by hydrogen and helium, but the remaining portion is dominated by oxygen, carbon, and nitrogen **(B)** Carbon, nitrogen, hydrogen, oxygen, sulfur, and phosphorus are distributed between five classes of important biomolecules. Nitrogen occurs in what are arguably the two most important - genetic information (nucleic acid) and the structural and catalytic molecules that interact to produce metabolism (proteins). **(C)** Simplified abiotic synthetic pathways of life's biochemical building blocks (adapted from Kitadai & Maruyama, 2018). **(D)** These fundamental building blocks are found in meteorites (shown in log scale). Sugars are found up to 180 parts per million (ppm) in the Murchison (carbonaceous chondrite, CM) meteorite (Furukawa et al., 2019). Amino acids, the most abundant, can be found up to 21 ppm within CM Chondrites and 2400 ppm within CR Chondrites (Glavin et al., 2010; Pizzarello & Shock, 2010). Fatty acids can be found up to 1000 ppm and 10 ppm within CM and CR Chondrites respectively (Lai et al., 2019). Nucleobases are found least abundantly up to 34 parts per billion (ppb) in the Murchison meteorite (Oba et al., 2022); nucleotides have never been detected within extraterrestrial material.

Looking beyond components of fundamental biochemistry, many other organics form under plausible prebiotic conditions. Some, such as sulfonic and hydroxy acids, are fully capable of forming polymers (Guo & McGrath, 2012; Frenkle-Pinter et al., 2022). While the functional potential of polymers made from these alternatives is underexplored, especially their potential to form catalytic enzyme analogs, amino acids already show some unexpected advantages to an origin of life. The esters that connect hydroxy acids, and the thioesters which connect sulfonic acids are, for example, less stable to hydrolysis than the peptide bonds which link amino acids (Robinson & Tester, 1990). This difference in stability contributes to a self-purification of depsipeptides (heteropolymers comprising a mixture of amino acids and hydroxy acids) in an environment that cycles through wet and dry conditions (Fig. 2; Forsythe et al., 2015). The most likely abiotic (hetero)polymers show the potential to develop into peptide sequences during chemical evolution. Spontaneous self-purification towards amino acid enrichment of depsipeptides is particularly relevant to extraterrestrial life because it suggests how even different starting points for polymer-based catalysis might converge upon amino acids over time.

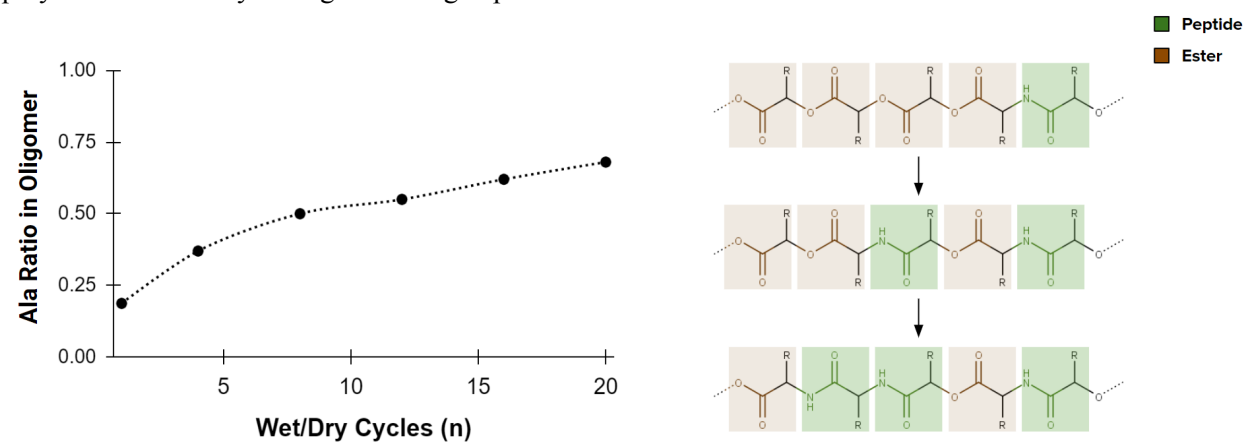

**Figure 2.** Heteropolymers comprising a mixture of amino acids and hydroxy acids (depsipeptides), exposed to wet-dry cycling, become enriched in amino acids (adapted from Forsythe et al., 2015). This enrichment, in part, is due to the stability difference between peptide (C-N) and ester (C-O-C) bonds suggesting the eventual convergence of an amino acid homopolymer (peptide) over time.

In summary, amino acids emulate the same properties that motivate current explorations for extraterrestrial life to "*follow the water*" (Irion, 2002, Schwieterman et al., 2018). While we cannot rule out life-sustaining possibilities of other solvents such as methane (McKay & Smith, 2005) or supercritical carbon dioxide (Budisa & Schulze-Makuch, 2014), the unique biophysics of water (e.g. Lynden-Bell et al., 2010; Finney, 2004) combines with its cosmic abundance (Mottl et al., 2007) to justify its central role within biochemistry. The same themes of ready availability and unusually useful physicochemical properties are true of amino acids: they form almost unstoppably within the abiotic organic chemistry that occurs in the presence of liquid water (e.g. Aponte et al., 2020). They can spontaneously enrich within heteropolymer sequences (e.g. Forsythe et al., 2015). Once polymerized, they display an amazing versatility to perform catalytic and structural roles in environments which overlap significantly with those identified for extraterrestrial environments (e.g. Merino et al., 2019). Thus, while certainties elude any scientific inquiry that looks beyond biochemistry as we know it, amino acids are excellent candidates with which to extend logically the current search for extraterrestrial life around water.

**Table 1.** Amino Acids are smaller than Other Components of Life's Biochemistry.

| | Heavy Atoms | Molecular Weight (g/mol) | Chemical Elements |
|---|---|---|---|
| **Coded Amino Acids** (ACDEFGHIKLMNPQRSTVWY) | 5-15 | 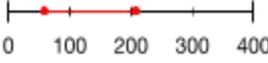 | CHONS |
| **'Prebiotic' Amino Acids** (ADEGILPSTV) | 5-8 | 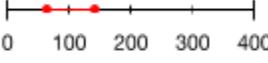 | CHONS |
| **Nucleobases** | 8-11 | 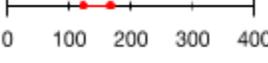 | CHONP |
| **Nucleotides** *Nucleobase + Ribose + PO₄* | 23-24 | 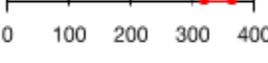 | CHONP |
| **Fatty Acids** | ≥5 | 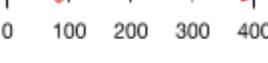 | CHO |
| Propionic acid§ | 5 | 74 | |
| Decanoic acid† | 12 | 172 | |
| **Lipids** | ≥12 | 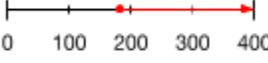 | CHO |
| Triformin* | 12 | 176 | |
| **Sugars** *Monosaccharides* | ≥6 | 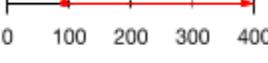 | CHO |
| Triose‡ | 6 | 90 | |
| Ribose | 10 | 150 | |

*Smallest triglyceride; ‡Smallest monosaccharide; †Prebiotically plausible membrane former (Todd et al., 2022); §Smallest fatty acid

# Would a xeno biochemistry use monosubstituted L-α-amino acids?

## L- vs D- stereochemistry

Were a xeno biochemistry to use amino acids, innumerable options exist quite different from life as we know it. In terms of chemical nomenclature, the genetically encoded alphabet is dominated by L-α-amino acids, where "L" and "α" each denote subsets of a far larger set of amino acid structures. A useful next level of inquiry can therefore ask why the subunits for genetically encoded proteins of life on Earth are restricted to (i) L-enantiomers of (ii) $\alpha$-amino acids.

**Would a xeno-biochemistry use L-amino acids?** The designation "L-" specifies one of the two mirror-image conformations possible for a single side-chain bound to a single ($\alpha$) carbon atom situated between the C- and N- termini (Box 1; Fig. 3). Because these termini are different, the two positions at which a side chain can bond to the intervening carbon atom form different (non-superimposable) three-dimensional molecules: L- versus D- enantiomers. 19 of the 20 genetically encoded amino acids are L-amino acids. The 20th, Glycine, is a unique exception to this pattern. In addition to the single hydrogen present on all 20 (monosubstituted - see below) amino acids, glycine has a second hydrogen atom attached to the alpha-carbon whereas all others have a side chain (R- group). From the perspective of the set used to construct genetically encoded proteins, Glycine is therefore most usefully perceived as the point-of-origin (zero) for the L-series, or indeed any series: effectively the absence of a side chain.

Without catalysts, such as protein enzymes, undirected chemical syntheses of amino acids generally produce equal amounts of L- and D- enantiomers. Excepting some rare reports of L-enantiomeric excess (Elisa et al., 2016), this racemic mixture is exactly what is observed as the norm in meteorites (Cronin & Pizzarello, 1983; Elsila et al., 2016) and the results of prebiotic simulation experiments (Strecker, 1850; Ashe et al., 2019; reviewed in Masamba, 2021). D-amino acids are, furthermore, synthesized and used throughout contemporary biology (reviewed in Sasabe & Suzuki, 2019; Grishin et al., 2020). The venom of the desert grass spider *Agelenopsis aperta*, for example, is known to contain D-serine at position 46 in omega-agatoxin IVB (Heck et al., 1994) while D-Serine functions as a signaling molecule mediating NMDA receptor activity in mammalian brains (Wolosker et al., 2008). Biological use of D-amino acids does not extend, however, to genetic decoding. D-enantiomers are generally toxic to cells to the degree that many prokaryotes and eukaryotes have evolved detoxification enzymes which control their concentration within the cell, sometimes by converting D-amino acids into L-equivalents (Pollegioni et al., 2007; D'Aniello et al., 1993). A common cause of this toxicity is that D-amino acids interfere with "normal" (ribosomal) genetic decoding and organisms which incorporate D-amino acids into peptides usually do so through specialized, "*non-ribosomal peptide synthesis.*" (Maruyama & Hamano, 2023). For example, D-alanine and D-glutamate are incorporated into cell wall structure by all bacterial cell walls that are known to contain peptidoglycan (Vollmer et al., 2008). But the ribosome, and thus its tolerance for the molecules into which it translates genetic messages, is a product of biological evolution. We may conclude that D-enantiomers were both available to life's origins and throughout its subsequent evolution: for post-LUCA genetic decoding, they are selected against for similar reasons that would disadvantage a UK motorist attempting to drive on the left side of roads in the USA. But was there any reason why the L-enantiomer was the direction in which this situation would resolve if played out a second time by another instance of biological evolution?

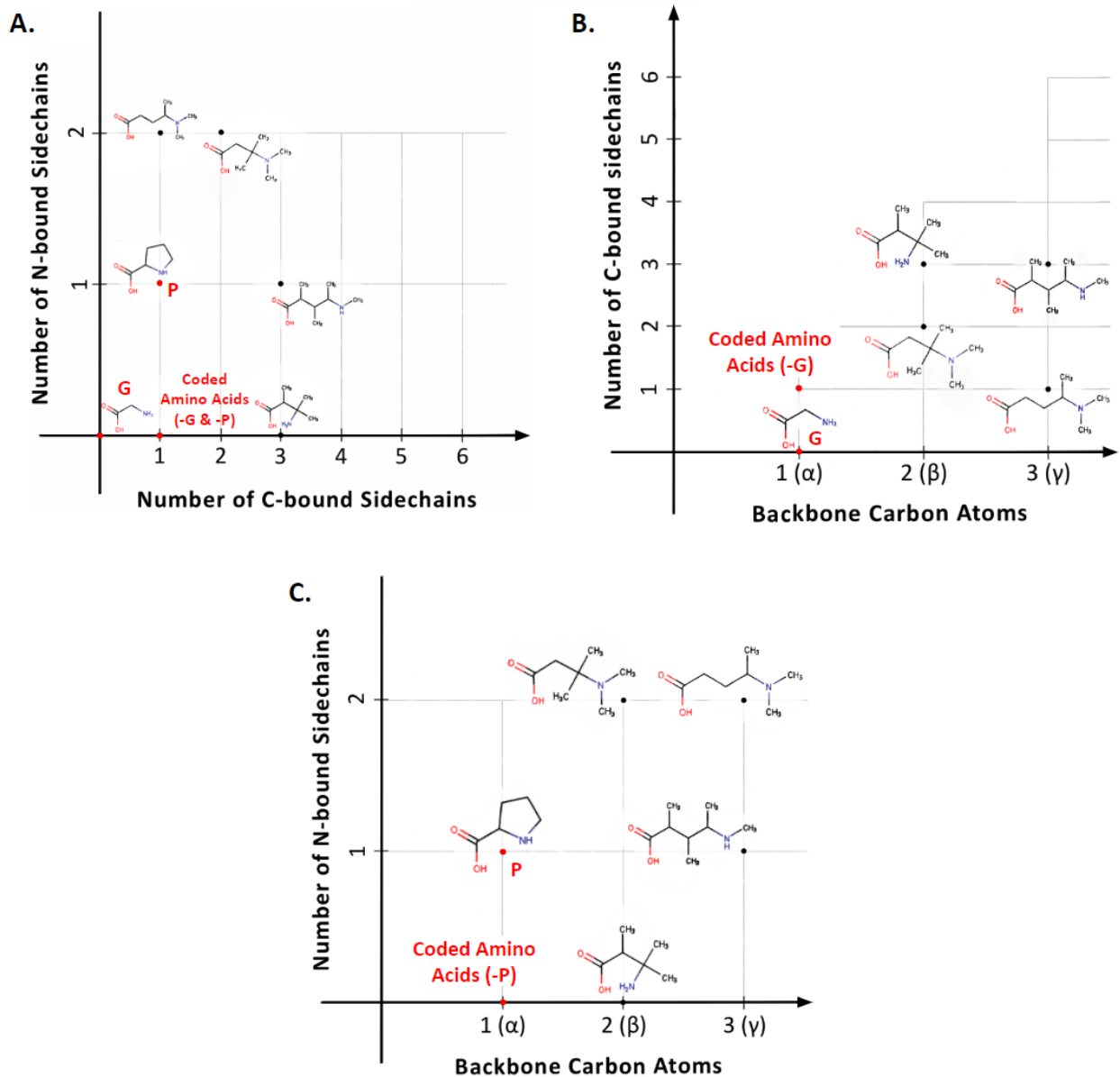

**Figure 3.** The universe of amino acid structures. **(A)** Distribution of amino acids based on the number of C-bound side chains vs N-bound side chains (genetically encoded amino acids highlighted: Ala, Pro, Gly). **(B)** With each additional C- atom in the backbone, the number of possible C- bound side chain attachment sites increases by 2. The coded amino acids, except glycine, are merely a point in this possible space. **(C)** While C-bound side chain attachment sites are theoretically infinite, the backbone nitrogen can only attach two side chains while retaining its neutral valence. Here, the genetically encoded amino acids, except proline, exist in simply one point within the possible space.

Alluded to above, one possible answer involves "*nonbiological enantiomeric enrichment processes prior to the emergence of life*" (Elisa et al., 2016). The bias reported from some simulations of interstellar and circumstellar astrochemistry illustrates this (Evans et al., 2012). However, the causal mechanism(s) for this or any other source of L-enantiomeric bias remains unclear. The literature of suggestions inspired by physics alone includes multiple ideas for direct molecular interactions with light (Davankov, 2009; Jorissen & Cerf, 2002; Ozturk & Sasselov, 2022; Kawasaki et al., 2005) and equally diverse explication of the long-recognized (Pasteur, 1860) role for crystal formation (Weissbuch & Lahav, 2011). While both families of explanation could perhaps merge into one, more general set of ideas relating to symmetry-breaking (Noorduin et al., 2009; Takahashi & Kobayashi, 2019 ), shifting perspective to chemistry finds an equally diverse set of competing and overlapping suggestions to explain enantioenrichment as a result of reaction pathways instead (e.g. Formose, Breslow & Chen, 2010; Strecker, Wagner et al., 2017). Thus, until further evidence resolves the current lack of consensus (reviewed in Toxvaerd, 2009; Percec & Leowanawat, 2011; Blackmond, 2019), it is perhaps more helpful to notice where the vying explanations agree: all tend to involve L-enantiomer *enrichment*, a bias, rather than complete absence of the D-enantiomer. This in turn implies that life's homochirality arose through some sort of evolutionary feedback, whether physical, chemical (Bryliakov, 2020) or, shifting to a third perspective, biological (e.g. Ruiz-Mirazo et al., 2014; Preiner et al., 2020). Viewed in this light, the unknown variable at present is the *extent* of evolutionary feedback versus a foundational bias laid by physics and/or chemistry (Bonner, 1991; Sallembien et al., 2022).

Recognizing a role for evolutionary feedback also provides reasons to retreat from rigid answers that once seemed clear regarding sub-questions within the topic of "L- versus D-". For example, beneath the question of which amino acid enantiomer is genetically encoded lies the simpler question: why are the genetically encoded amino acids homochiral rather than heterochiral? Cleaves (2010) summarizes a longstanding view (e.g. Brack & Spach, 1979; Nanda et al., 2007) that "*the exclusive use of one isomer allows for the formation of regular secondary structural motifs.*" The idea here is that helices and sheets, the foundations of a 3-dimensional protein structure, are stabilized through intramolecular interactions that would be obstructed by a haphazard mixture of L- and D- amino acids (Fig. 4). However, recent evidence shows that heterochiral proteins, "*though less stable than their homochiral analogues, exhibit structural requirements (folding, substrate binding and active sites) suitable for promoting early metabolism*" (Sallembien et al., 2022: see also Weil-Ktorza et al., 2023). Thus, a better explanation than a biophysical necessity, that life can only originate using homochiral building blocks, is that evolution favored a homochiral set of amino acids for efficiency much as increasing size and speed of traffic caused nations around the world to decide which side of the road travels in which direction. In this sense, it seems clearer why replicating systems would be drawn to homochiral amino acids as monomeric building blocks than why it would use the L- versus D- enantiomers. Consensus wisdom has long held that "mirror-life" would function perfectly well (Pasteur, 1860; Wang et al., 2019; Fan et al., 2021) and, at an extreme, evolutionary competition could have led to the eradication of a fully functional 'D-life' (Green & Jain, 2010).

## Monosubstitution

A continuing focus on evolutionary feedback rather than biophysical necessity addresses a subtly different sub-question implied by L-homochirality: why do all genetically encoded amino acids use only one of the two possible side chain attachment points presented by the alpha carbon (i.e. why are they monosubstituted)? Like a mixture of L- and D- chiralities, the presence of two side chains on the alpha carbon ($\alpha,\alpha$-disubstituted) has long been recognized to obstruct secondary structure formation (e.g. Fretheim et al., 1979; Fig 4B). However, while "*the secondary structure of [disubstituted amino acid] peptides [are] especially restricted,*" (Tanaka, 2007), subsequent evidence again shows that structure formation is possible (Wang et al., 2022). A more robust explanation for a monosubstituted alphabet is what McKay (2004) calls the Lego Principle: "*Biological processes, in contrast to abiotic mechanisms, do* *not make use of the range of possible organic molecules. Instead, biology is built from a selected set…General arguments of thermodynamic efficiency…suggest that this selectivity is required for biological function and is a general result of natural selection."*

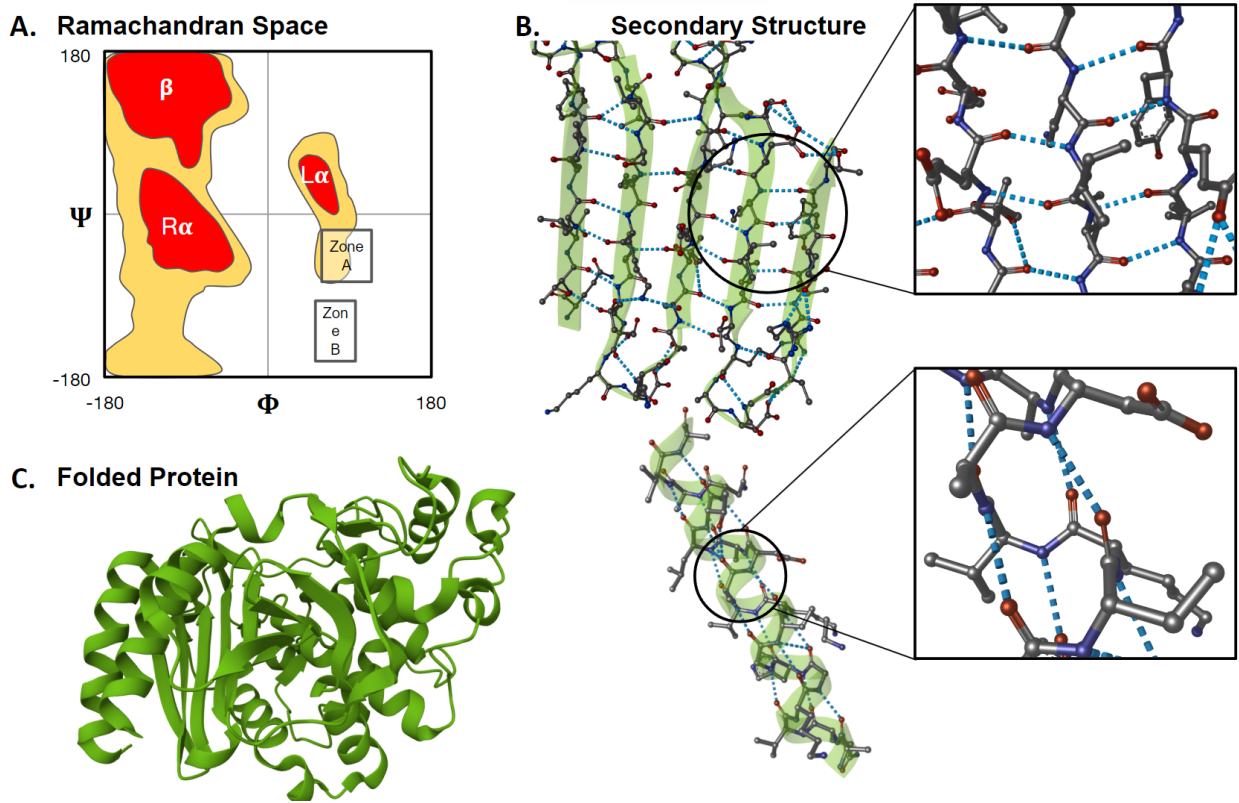

**Figure 4.** Xeno amino acids potentially change the biochemistry of protein folding **(A)** Combinations of peptide torsion angles (*ϕ* & *Ψ*: see Box 1) summarize peptide/protein secondary structure (adapted from Elsliger et al., 2012). Zones A & B are torsion-angle pairs disallowed by biophysics under strict steric considerations yet observed empirically upon careful investigation for specific combinations of side chains (discussed in Kalamankar et al., 2015). This provides one clue that the current, well-established map of secondary structure could shift or become unrecognizable if rebuilt for xeno amino acids **(B)** The dominant secondary structures of life on Earth, α-helices and ☐-sheets (first described by Linderstrøm-Lang, 1952), form and are stabilized by hydrogen bonds (blue dotted lines). These hydrogen bonds usually involve atoms in the amino acid backbone. The bonds and therefore protein structures would be obstructed or altered if peptides comprised a heterogeneous enantiomeric mixture, more than one sidechain per amino acid or longer backbones (i.e. β- γ- δ-) **(C)** Multiple secondary structures within a single polymerized amino acid sequence combine to form the larger, folded tertiary structure. It follows from A-C that protein structures could be unpredictably different if the fundamental building blocks (amino acids) were changed. *Images were created using Mol\* Viewer (Sehna et al., 2021) with PDB 4LV0 AmpC beta-lactamase in complex with m-aminophenyl boronic acid (London et al., 2014).*

### *α*-amino acids versus longer backbones

Viewing amino acids as building blocks for biopolymers offers further explanatory power to turn from "L-" to "α-" as a feature of genetically encoded amino acids. Alpha amino acids are those in which a single carbon atom is situated between the C- and N- termini (Box 1). However, the number of carbon atoms here can be larger, with *α*-amino acids as the simplest structural subclass within a theoretically infinite series: β-, γ-, δ- etc. (Fig 3). Each carbon atom added to the backbone adds two, different positions at which a side chain could attach. Thus, whereas an α-carbon atom can be only mono- or di-substituted, the addition of a β-carbon atom permits up to four side chains, γ- up to six, and so on. The addition of each backbone carbon atom, a potential new chiral center, thus increases structural possibilities exponentially. For example, the 6 possible carbon side chain attachment sites in a γ-amino acid imply that two different side chains could occupy any of 30 ($_6P_2$) different permutations and 6 different side chains could be arranged in 720 different ways ($_6P_6$).

      Just as was the case for chiral alternatives, longer amino acid backbones are far more than a theoretical possibility. The same laboratory simulations and meteorite analyses that indicate the prebiotic plausibility of D- enantiomers and disubstituted α-amino acids also reveal the presence of β-, γ-, and δ- amino acids (Cleaves 2010: Table 1). Further echoing the situation described above for alternatives to the L-enantiomer, amino acids with additional backbone carbon atoms also occur throughout present-day biology (Lelais & Seebach, 2004). Once again, themes of cell signaling and allelochemistry (defensive and offensive toxicity) surface amidst a broad repertoire of functions. For example, the simplest γ-amino acid, gamma-aminobutyric acid (GABA), functions as both an important neurotransmitter within animals and a cell signaling molecule in plants (Li et al., 2021; Sigel & Steinmann, 2012) whereas β-Methylamino-L-alanine, or BMAA, is a powerful neurotoxin to mammals produced by cyanobacteria (Cox et al., 2005) and plants (Vega & Bell, 1967). Clearly, it is not beyond the reach of biological evolution to work with longer backbones or indeed a heterogeneous mixture of different backbone types.

      From the perspective of peptide folding, however, longer backbones produce less stable secondary structures. Each carbon-carbon bond within the amino acid backbone permits rotation that increases the flexibility of a peptide chain (Kiss et al., 2017; Legrand and Maillard, 2021; Nagata et al., 2023; also see Fig 3). But while the lack of backbone rigidity has long been noted as one, simple, reason why natural selection would favor α-amino acids for the genetic code (Weber and Miller, 1981), synthetic biology demonstrates that β-amino acid peptide structures are possible (Steer et al., 2002; Fülöp et al., 2006; Forsythe et al., 2018). Thus arguments for the exclusion of β-, γ-, and δ- amino acids from the genetic code again resemble those for homochirality: a more robust explanation than strict biophysical constraint is evolutionary optimization. If this is an evolutionary outcome then perhaps one might need to look no further than the higher energy cost for any cell working with a more diverse repertoire of building blocks. For example, to polymerize a mixture of "*α- and β- amino acids, four enzymes would probably be required. One enzyme would be needed for each of the four combinations of substrates: α, α; α, β; β, α and β, β*" (Weber & Miller, 1981). Even simpler, the additional carbons in the amino acid backbone would result in a more expensive metabolism because it would logically take more energy to synthesize, manipulate, move, and even degrade the more massive proteins that are built with β-, γ-, and δ- amino acids.

      A current understanding of why Earth's biology genetically encodes L-α-amino acids acknowledges that other options including D-enantiomers, disubstituted amino acids, and longer backbones were plausibly available to life's origin and evolution. Against this background, explanations for life as we know it have retained a central theme since their inception that there might be some advantage to alpha, monosubstituted amino acids for linear polymers which fold into complex 3-dimensional shapes. What has changed over time has been a retreat from "hard" statements about impossibility of other chemical structures in favor of "softer" statements about preferential attributes of those found in the canonical alphabet. Whereas comparisons between amino acids and other types of molecules (e.g. sulfonic or hydroxy acids) locate these advantages within physics (the strength of the

peptide bond), comparisons between different backbones add to this signs of evolutionary influence, and explanatory power for side chains currently favor biological evolution. For example, other backbone types are capable of forming polymers but would seem likely to cost more energy, both directly and indirectly. Thus, while no direct evidence shows that primordial life used another type of molecular backbone for structural and catalytic polymers (if anything, it would seem more likely that it could have used a mixture of different building blocks), increasingly clear reasons suggest why evolution would favor the streamlining of any such alphabet into the single type of repeating structure encountered within genetically encoded L-$\alpha$-amino acids: Whether or not life on Earth began with L-$\alpha$-amino acids, it can see why it would have evolved to this state. Since the reasoning involved comes from physics and chemistry, it would seem unsurprising to discover a similar outcome within an independent origin of life (a xeno biochemistry), so long as something like natural selection has caused replicating systems to distinguish themselves from the abiotic universe (Levin et al., 2017). To expect otherwise, science would need to identify specific physical conditions under which functional advantages of larger and/or more heterogeneous molecules would outweigh their cost.

## Would a xeno biochemistry use different **side-chains**?

The third and final amino acid attribute that deserves careful consideration for alternative biochemistries is the set of 20 side chains used within the standard genetic code. Both backbones and side chains contribute to producing protein structure, but differently so. Whereas it is the unvarying features of the L-$\alpha$-backbone that matter (notwithstanding the special cases of Glycine and Proline above), it is the differences between side chains that are important. In 1972, the Nobel Prize in Chemistry was awarded for finding that "*at least for a small globular protein in its standard physiological environment, the native structure is determined only by the protein's amino acid sequence*" (Anfinsen, 1973). This finding ended some fundamentally misled ideas about protein folding (e.g. cyclol hypothesis: Frank, 1936; see Pauling & Niemann 1939) with the knowledge that set of 20, genetically encoded side chains define what protein shapes and functions can be genetically encoded by life (Chiarabelli et al., 2006a; Chiarabelli et al., 2006b). Thus, while the "Lego" principle accounts neatly for an unvarying, L-$\alpha$-backbone (Fig. 4), any explanation of side chains must introduce new ideas to explain diversity.

Like the discussion of backbones presented above, plenty of plausible alternatives exist to the set of 20 side chains genetically encoded by life as we know it. Once again, these options are informed by prebiotic chemistry (both simulations and meteorite analysis; Fig. 5A), and by their widespread use within biology (Fig. 5B). Indeed, early glimpses of diversity (Wong and Bronskill 1979, Uy & Wold, 1977) reflected limitations of instrumentation more than chemical reality for both abiotic (Cleaves, 2010; Aponte et al., 2020) and biological amino acids (Flissi et al., 2019; Fekkes, 2012). Complementing these naturally occurring alternatives is plentiful experimental evidence that other side chains can still function within the genetic code, even after 3.5 billion years of evolution. In recent years, synthetic biologists have engineered more than 250 different amino acid side chains into protein synthesis (Mayer-Bacon et al., 2021). Indeed, the subfield of non-coded amino acids (ncAA's) is developing so fast that the total of 250 is out of date and any alternative suggested here would be obsolete within months (eg. Lee et al., 2019; Andrews et al., 2022). Such technological progress aligns well with the widespread use of xeno amino acids in specialized versions of peptide synthesis (Reimer et al., 2018; Dell et al., 2022) to suggest the imminent delivery of human-engineered alternative amino acid alphabets (Feldman et al., 2019).

Given clear evidence for a multiplicity of alternatives, it is useful to remember that side chain diversity directly defines the corresponding universe of shapes and functions. Indeed a major challenge for current research is the theoretically infinite diversity of side chains made possible by organic chemistry. Although imposing a maximum side chain size (e.g. by volume, number of atoms, etc.) constrains the set into a finite number, any such number is problematically large (Fig. 5D). There are, for example, approximately $5.6 \times 10^5$ isomers of the side chain for Tryptophan (Meringer et al., 2013), the largest of the coded amino acids by volume, before adding the cumulative side chains smaller than this, and/or those encompassed by slightly different atomic composition. Since synthetic biology has already successfully incorporated L-$\alpha$-amino acids far larger than Tryptophan, and far more chemically diverse than anything seen in the genetic code, into "natural" (ribosomal) genetic decoding (Fig. 5B), perhaps the single clearest idea for xeno side chains at present is that those used by post-LUCA life on Earth are not the only set of chemical structures capable of linking into functional biopolymers. The more interesting question is what shaped this particular evolutionary outcome?

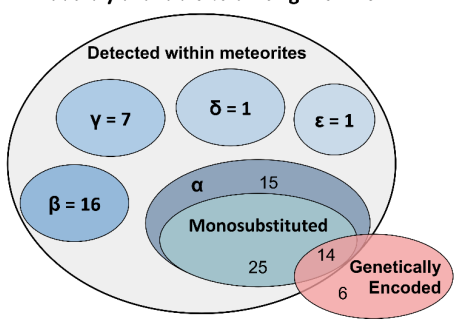

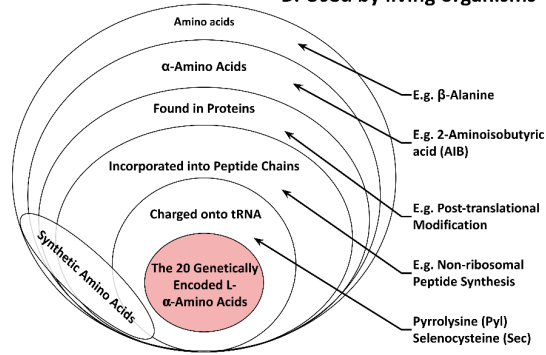

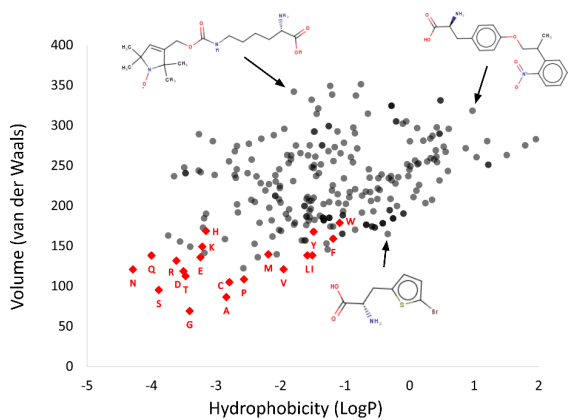

**Figure 5:** Like amino acid backbones, many side chains are possible beyond the 20 found within the standard genetic code. **(A)** abiotic synthesis: various other side chains are found in meteorites and produced by prebiotic simulations (see Mayer Bacon et al. 2022, supplementary data); **(B)** biological: many other side chains are used by living organisms (see Freeland, 2009); **(C)** synthetic biology has successfully incorporated hundreds of alternative side chains into protein synthesis, including numerous far larger than anything found within the standard genetic code (Lui & Schultz, 2010; Dumas et al., 2015; Noodling et al., 2019; also see Mayer-Bacon et al., 2021). **(D)** theoretical: the addition of each carbon atom increases exponentially (combinatorially) the number of chemical structures that are possible (see: Meringer et al., 2013).

Given amino acid side chains' importance in defining protein structure and the clear potential for alternatives, surprisingly little research has addressed the consequences of building proteins with other side chains. The initial success of prebiotic simulations and their alignment with meteorite analyses certainly inspired a small, early sub-literature considering amino acids from beyond the genetic code, usually in the form of a "deep dive" into one particular amino acid (e.g. Anfinsen and Corley 1969 for Norleucine; Jukes, 1973 for Ornithine, but see also Weber and Miller 1981 for a review). However, in the later years of the 20th century, focus narrowed to how the 20 amino acid "meanings" became incorporated into the standard genetic code rather than looking beyond. Certainly, to explore amino acids from beyond the genetic code is costly in terms of both time and money but considerable time and money were spent investigating the 20 (see AAIndex: Kawashima et al., 2008): so why did this activity not look beyond the molecules of the central dogma? No single reason clearly explains why but, with hindsight, several contributing factors may be inferred.

From the perspective of biology, the discovery that side chains steer protein folding emerged within a larger framework, represented by 4 other Nobel prizes (Beadle et al., 1958; Holley et al., 1968; Crick et al., 1962; Delbrück et al., 1969). Together, the work rewarded by these prizes describes how all life on Earth converts genetic information into protein-based metabolism. Within this "central dogma of molecular biology" (Crick, 1970), amino acids build biological proteins because they are programmed to do so by genetic information. A considerable literature thus developed to discuss how 20 amino acids became assigned to 64 different genetic code words (codons) (see, for example, Crick, 1966; Wong, 1975; Wong & Bronskill, 1979; Knight et al., 1999; Koonin and Novozhilov, 2017). Indeed, the genetic code remains central to biology such that in 2022 alone PubMed reported 64 new publications using the keyword terms *genetic code* and *evolution*.

From the perspective of prebiotic chemistry, the syntheses that accounted so easily for some of the amino acids found within the standard genetic code gave way to unexpected difficulties in accounting for the rest (Wong & Bronskill, 1979). Synthetic, organic chemists researching life's origins thus diverted efforts towards accounting for missing members of the 20 rather than exploring side chains that lie beyond.

From the perspective of protein structural biochemistry, the satisfyingly simple insight that side chains steer protein folding proved frustratingly difficult to model or predict with detailed physicochemistry. Levinthal (1968) famously captured the essence of the problem by pointing out the overwhelming number of possible conformations into which polymers built from an alphabet of 20 different side chains could potentially fold. Simple, pragmatic urgency of making progress in solving the "protein folding problem" (Kryshtafovych et al., 2005) replaced asking equivalent questions about other possible side chains, and another relevant research community focused on the 20 rather than looking beyond.

Finally, from the perspective of "origins" research, a sixth Nobel Prize awarded for the discovery of catalytic RNA (Altman & Cech, 1989). This extension of the Central Dogma led directly to the declaration of the RNA world hypothesis (Gilbert, 1986; Joyce, 2002; Štorchová et al., 1994) which was taken by many to imply that amino acids entered an evolved, RNA-based biology (e.g. Benner et al., 1989; Freeland et al., 1999; Yarus, 2017; Rivas & Fox, 2023). Under such thinking, the set of amino acids found within the genetic code is one that can be synthesized by metabolism rather than made available by prebiotic chemistry. The resulting shift in perspective is seen by comparing two influential review articles, separated by three decades. Whereas Weber and Miller (1981) used expertise in prebiotic chemistry to discuss which amino acid side chains would have been available to life's origin, by 2017 Doig explained that "*If protein synthesis arose from the RNA World… life was already biochemically sophisticated and the environment was substantially modified from the conditions prevailing during abiogenesis. Arguments based on prebiotic conditions are thus not especially helpful in rationalizing amino acid selection.*" While interpretations of the RNA-world hypothesis continue to diversify (Fine & Pearlman, 2023), the idea that the 20 genetically encoded amino acids reflect the evolutionary expansion of simpler, earlier code continues to gain multidisciplinary consensus (Fig. 6).

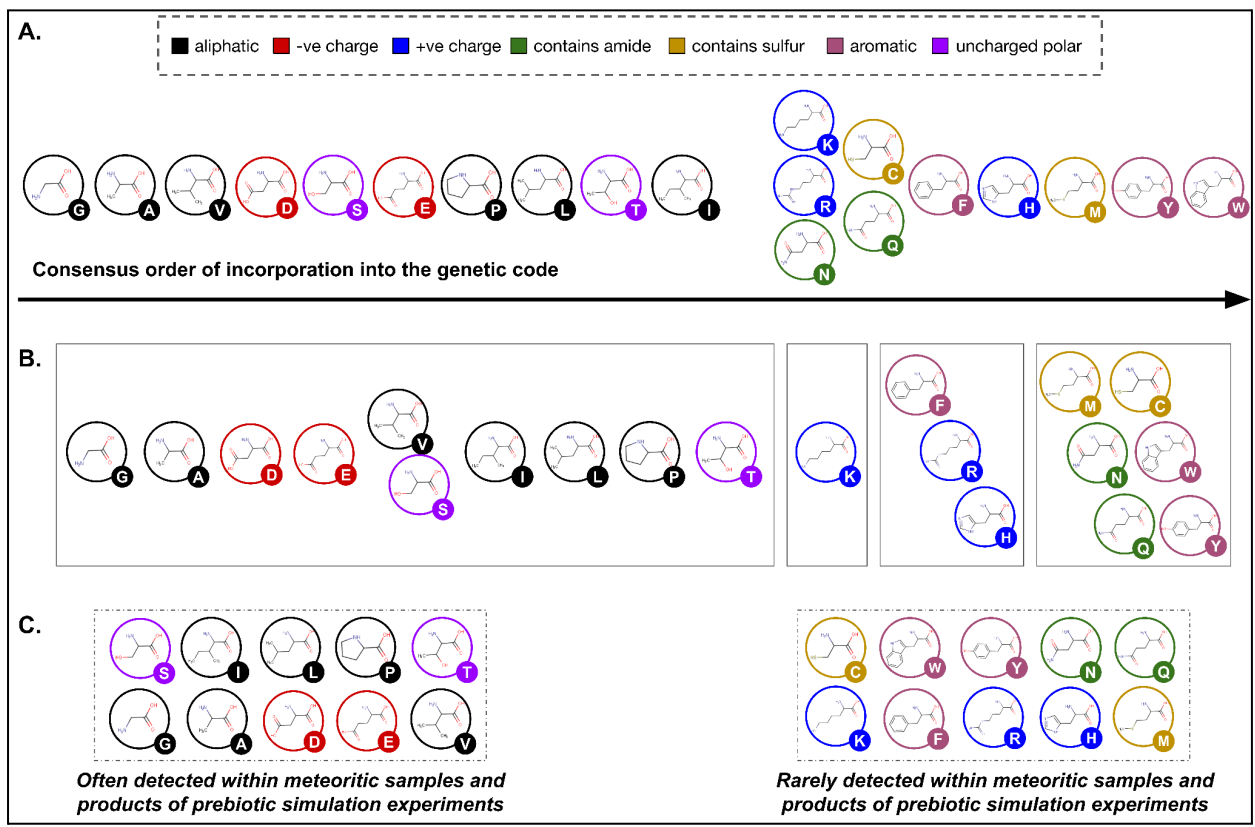

**Figure 6:** A comparison of three major syntheses of scientific literature concerning the antiquity of amino acids within the standard genetic code. All agree that the canonical alphabet of 20 amino acids evolved from an earlier genetic code involving fewer amino acids. **(A)** Trifonov (2000) analyzed 40 peer-reviewed publications about the evolution of the genetic code to calculate a detailed chronology by which the set of 20 became established. **(B)** Higgs & Pudritz (2009) analyzed a similar amount of different literature to arrive at broadly similar conclusions. **(C)** Cleaves (2010) focused on the literature of prebiotic chemistry alone (meteorites, spark tube experiments, and HCN polymerization) to agree with both. Adapted from Mayer Bacon et al. 2022.

Jumping ahead to the 21st century, the past decade has witnessed a resurgence of interest in looking beyond the genetically encoded alphabet of 20 side chains. At present, around ~25 peer-reviewed publications contribute directly to this literature (Fig. 7) and can be understood as deriving from three distinct research communities, each now equally relevant to xeno biochemistry: ***De Novo* Protein Design, Prebiotic Chemistry**, and **Molecular Evolutionary Biology**. In general terms, these different communities are worth distinguishing because, prior to a shared mutual interest in xeno amino acid side chains, their approaches connect only by going back further to foundational authors who wrote with great influence around the discovery of the central dogma of molecular biology. It is then the emerging, new synthesis of these three subfields which promises exciting new progress.

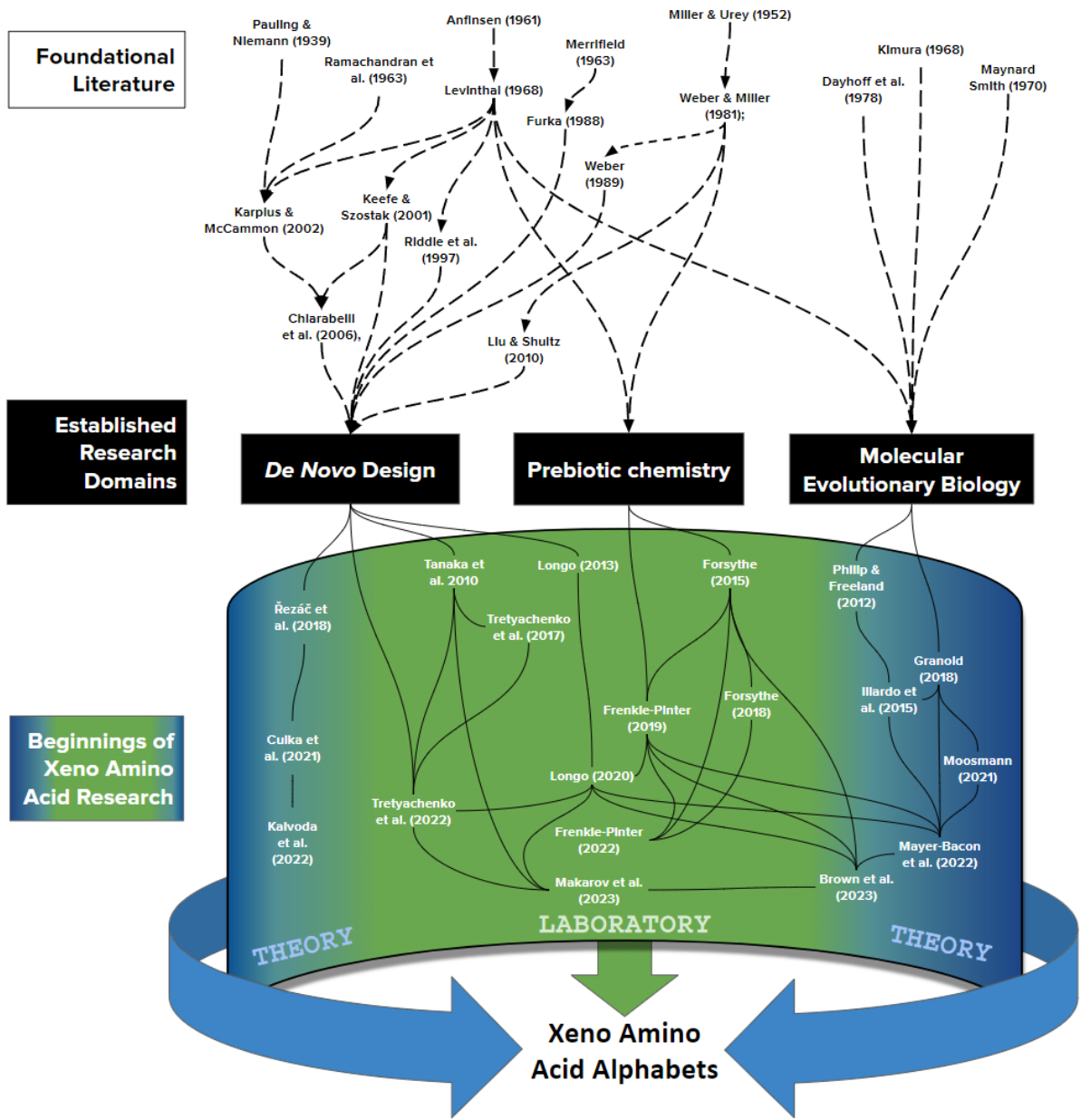

**Figure 7:** A literature map of recent (>2010) publications converging on direct exploration of alternative amino acid alphabets, traced back through diverse subdisciplines to foundational works. This literature involves both theory and experiment. Whereas experimental work is already starting to integrate the 3-named subfields (*de novo* protein design, prebiotic chemistry, and molecular evolution), relevant theory is at present siloed between two largely unconnected edges: sophisticated subatomic modeling of alternative side chains and biologically inspired design of xeno alphabets.

# Clues from *de novo* Protein Design: altering the functional units of life as we know it

*De novo* protein design builds from Anfinsen's (1961) demonstration that a protein's 3-dimensional structure is produced by the specific sequence of amino acid side chains. Woolfson (2021) characterizes three approaches that inform both the protein folding problem and provide a powerful foundation for adapting to xeno alphabet thinking.

(i) **Amino acid alphabet simplification:** It has long been speculated that an alphabet comprising fewer than 20 amino acids could build functional proteins (e.g. Brack and Orgel, 1975), and research working with reduced amino acid alphabets, or Minimal Protein Design, "*uses straightforward chemical principles such as patterning of polar (p) and hydrophobic (h) amino-acid residues to direct the folding and assembly of secondary structures…*" (Woolfson, 2021)**.** Riddle (1997) first demonstrated empirically that a random sequence of amino acids drawn from a reduced subset of that canonical twenty can exhibit structure and function. Tanaka (2010) then scaled up this observation by comparing 3 peptide libraries constructed from random sequences using alphabets of different lengths. Within this methodological framework, Longo et al. (2013) then built a "*foldable halophilic protein*" from an alphabet primarily reduced to those which are prebiotically plausible. From here it was a tractable and clear step to introduce xeno side chains (Longo et al., 2020). Tretyachenko et al. (2022) further advanced this approach by introducing high-throughput sequencing. Interestingly as the composition of prebiotically plausible amino acids increased to 100%, new folding principles started to emerge (Tretyachenko et al., 2022). Most recently of all, Makarov et al. (2023) has started to introduce xeno amino acid side chains within a reduced alphabet framework in order to compare canonical versus non-canonical side chains. Presently it remains to be seen just how many and how deep are the new protein-folding principles that come from building with xeno side chains.

(ii) **Rational peptide design** constructs peptides that sample a targeted region of protein sequence space using "*sequence-to-structure relationships garnered from biochemical, bioinformatics or empirical studies*" (Woolfson, 2021). The invention of solid-state synthesis (Merrifield, 1986) permitted researchers for the first time to synthesize protein sequences efficiently without involving life's molecular machinery for genetic decoding: a freedom powerful enough to earn yet another Nobel prize in chemistry (Merrifield, 1984). Furka's subsequent (1991) addition of a "mix and split strategy" added the power of combinatorial chemistry to this approach (for a recent review see Furka, 2022). The underlying and significant advantage here for studying xeno amino acids is a lack of dependence on life as we know it. This potential is, however, only now being realized (e.g. Makarov et al., 2023), and is currently limited to small oligopeptides. Indeed, no one has yet worked with an entirely xeno amino acid alphabet. The challenge is as much mathematical as biochemical: increasing the length of a peptide by each amino acid increases exponentially the possible sequence space. A peptide of length 100, for example, drawn from an alphabet of 20 amino acids can be any of $\sim 10^{130}$ possible sequences, enormously more than the number of atoms estimated to comprise the entire universe.

(iii) **Computational modeling** uses biophysics to understand protein folding *in silico* by generating and evaluating "*full atomistic models for many different sequences for a given design target … ahead of experimental studies*" (Woolfson, 2021). The "protein folding problem" was born when it was noticed that physics somehow sorts through pragmatically infinite conformational possibilities to produce Anfinsen's lauded outcomes ("Levinthal's Paradox," 1969). Progress in understanding how biophysics does so came in 1994 when the Critical Assessment of Protein Structure Prediction (CASP) coalesced diverse approaches into an annual competition. By 2005, the overview of results was that "*current major challenges are refining comparative models [as they fast approach] experimental accuracy.*" In other words, all the best protein fold prediction algorithms share one idea in common: begin by finding a protein structure already known to science that is similar to the one under current scrutiny. In contrast, predictions built using first principles of physics and chemistry were relegated to "*handle parts of comparative models not available from a template*" (Moult, 2005). Hope for merging these two approaches came from Rosetta (Simons, 1999) which showed that accurate protein structure (~300 amino acids) can be predicted from concatenated peptide fragments 5-amino acids long (5-mers), it was a return to homology modeling when DeepMind's AlphaFold 2 (Jumper et al., 2021) effectively solved the protein folding problem for most natural proteins, but only by taking the basis for its predictions into a black box.

      To know how xeno alphabets will fold, it is logically necessary to take the advice from CASP and focus on what we understand about the physics of "standard" protein folding, however challenging. Here, molecular dynamics (reviewed in Karplus & McCammon, 2002) paved the way for thinking about biomolecules in terms of "conformational dynamics" (McCammon et al., 1977), which means comparing the Gibbs free energy of different possible 3-dimensional conformations as a guide to understanding the stable one which biological polymers find. This "*ab initio*" approach still cannot predict protein structure with anything like the power and accuracy of AlphaFold 2, but recent advances in quantum mechanical models (as seen in Culka et al., 2021; Kalvoda et al., 2023) are fast building the framework for future prediction of oligopeptides incorporating, or entirely comprising of xeno amino acids.

      To Wolfson's three subfields, a careful discussion of xeno amino acids may usefully add a fourth: **Alternative genetic codes.** A central point of the central dogma is that life decodes polymerized sequences of nucleotides (genes) into polymerized sequences of amino acids (proteins) (Box 1) and yet it has been traditional to study one type of biopolymer or the other. RNA and protein research have populated the pages of different journals, the authors of these two literatures have gathered at different conferences, developed different specialized terminologies, and generally checked many boxes for being considered as different academic disciplines (Krishnan 2009). Wolfson, a distinguished protein researcher writing for the protein community, focuses on amino acid alphabet simplification, rational peptide design, and computational protein modeling. Alternative genetic codes add something different simply by exploring protein structure and functions from the perspective of genetics, where a host of tools were developed in the wake of the central dogma to study and manipulate genetic material. For example, when Keefe & Szostak (2001) first explored the frequency at which folds and functions occur within protein sequence space using (mRNA) phage display, their artificial selection of RNA sequences under controlled mutation rates deliberately emulated the power of natural selection to find new, functional proteins. Twenty years later this approach has matured to bring into view, among much else, user-defined ("programmable") genetic codes (Chin, 2011; de la Torre & Chin, 2020). This potential meets impressive progress by synthetic biologists who have engineered more than 250 different amino acid side chains into protein synthesis (Mayer-Bacon et al., 2021). At least some of this effort is with an eye toward systems in which "*... a [semi-synthetic organism] is now, for the first time, able to efficiently produce proteins containing multiple, proximal ncAAs*" Feldman et al. (2019).

## Clues from Prebiotic Chemistry: bridging the gap between life and the non-living universe

From the perspective of prebiotic chemistry, the two parallel and intertwined perspectives with which we introduced this review remain directly relevant to understanding xeno amino acids: chemical simulations in the laboratory (Bada & Cleaves, 2015) and direct analysis of prebiotic environments (Aponte et al., 2020).

Instead of working forward from prebiotic chemistry, others have tried to work backward from the post-LUCA genetic code. Jukes (1973) was among the first to provide a specific candidate sidechain (ornithine) as a possible forerunner to the genetically encoded amino acid arginine. Additional contributions over the next couple of decades were relatively sparse and led a thorough review to conclude their probable irrelevance for reasons of biophysics: ornithine peptides, for example, "*are unstable because internal lactamization*" (Weber & Miller, 1981), where lactamization refers to a carbon-bound, linear, side chain bending around to also bond with the backbone amine, producing a cyclical structure in the subclass of amino acids to which proline belongs (Figure 3B). In this sense it was the growth of the RNA world hypothesis that clarified "*Arguments based on prebiotic conditions [alone] are thus not especially helpful in rationalizing amino acid selection.*" (Doig, 2017). In other words, once evolution by natural selection is at work on the alphabet, it is fully capable of introducing amino acids that are prebiotically implausible. Moving into the 21st century, growing acceptance of the idea that terrestrial life genetically encodes up to 22 amino acids (Atkins & Gesteland, 2002) provides another kind of empirical evidence that the genetically encoded alphabet can and does evolve. Wong & Bronskill (1979) first overtly introduced the idea that an explanation for the standard alphabet would have to include both prebiotic chemistry and subsequent biological evolution. And yet, until the 21st century, the 20 still dominated all knowledge about amino acids in proteins.

Recent work has begun to explore xeno side chains in comparison with canonical amino acids. Fenkle-Pinter et al. (2019), for example, have concluded that canonical side chains seem predisposed to form polymers more readily than non-canonical alternatives on biophysical grounds. By comparing the readiness with which lysine, arginine, and histidine form peptide bonds with each other, as opposed to with analogs from beyond the standard alphabet (ornithine, 2,4-diaminobutyric acid, and 2,3-diaminopropionic acid), this work concludes that "*the proteinaceous amino acids exhibit more selective oligomerization [suggesting] a chemical basis for the selection of Lys, Arg, and His over other cationic amino acids.*" While the results and indeed the question are pioneering important new information, it is for now puzzling that the particular amino acids studied are ones which other disciplines, from meteoritics to molecular evolution, agree entered the code only after enzyme-based metabolism had removed any semblance of proteins forming by competing to oligomerize (Fig. 6).

## Clues from Molecular Evolutionary Biology: natural selection guiding alphabet design

From the perspective of evolutionary biology, the discovery of life's central dogma defined for the first time specific, universal parameters with which life evolves at a molecular level. This replaced generations of mathematical modeling built creatively and cleverly on the limited knowledge that genes are particulate (non-blending), occur on chromosomes, and mutations change them: foundational rules that were famously criticized as "*bean bag [population] genetics*" (see Rao & Nanjundiah, 2011 for a review).

In the aftermath of the central dogma, three pioneers of evolutionary theory developed the new potential for molecular detail from different directions. Motoo Kimura (1968) used the new molecular knowledge to translate older, powerful population genetics that emphasized the role of chance (genetic drift) relative to natural selection. He noticed quickly that the central dogma implies significant "selectively neutral" evolution through, for example, redundancy in the genetic code (Jukes & Kimura, 1984). In contrast, Maynard Smith (1970) pioneered how to think clearly about natural selection at the molecular level, picturing adaptive walks through sequence space. Margaret Dayhoff (1978) complemented both approaches by using computing to summarize and then analyze empirical patterns of molecular evolution. She extracted quantitative statements about the patterns by which amino acids substituted for one another over time, largely corroborating Kimura's thinking as the strongest signal within molecular evolution.

The combined insights of these individuals and the work they subsequently inspired combined to move molecular evolution into a much more central role within biological and biomedical research. For example, bioinformatics is built around evolutionary ideas such as homology and phylogeny. It turns out these ideas are important for research that ranges from predicting protein folds to finding and understanding the role of protein-coding genes. However here, just as we noted more generally in the introduction for all of biology, the resulting focus on one, canonical alphabet of 20 possible amino acids has produced a contemporary science that is surprisingly blind to how molecular evolution would change if a different set of side chains were involved. It is, for example, easy to imagine why xeno amino acids could bring new physicochemistry of protein folding if we focus on the "standard" amino acid cysteine. A specific and unique characteristic of cysteine, one which contributed directly to Anfinsen's Nobel prize-winning work, is that two instances of cysteine at very different places within a single protein sequence can form disulfide bridges with one another as the protein folds into a 3-dimensional shape. If the genetic code lacked cysteine then nothing like disulfide bridges would exist among the other 19 to inform us (or a machine learning algorithm) of their possible existence and role in protein folding. Less extreme but more widespread than new covalent bonds, "*side-chain and backbone interactions [within 'natural' protein sequences] may provide the energetic compensation necessary for populating [hitherto unrecognized] region of $\varphi$–$\psi$ space*" (Kalamankar et al., 2014). If sidechain physicochemistry of the 20 can still expand understanding of sequence/structure relationships, then it would seem unwise to expect an indefinitely large and diverse set of xeno side chains not to alter these relationships further.

Almost everything we know about each member of the canonical alphabet is relative to the other 19. The challenge is knowing where to focus within a vast set of possible side chains (see Meringer et al., 2013) and an equally vast set of possible biophysical properties (e.g. Ghosh et al., 2023)

One sustained attempt to circumvent this limitation draws inspiration from the evolutionary methodology of optimality theory (Parker & Maynard Smith, 1990). When we wish to understand an aspect of the living world, we may ask what about it is unusual and plausibly the result of natural selection? Doing so quantitatively, in the context of plausible alternatives, begins a framework for scientific exploration of evolutionary cause(s). The idea is not that the initial hypothesis is correct, but rather it is a way for the researcher to enter an iterative cycle of comparing predictions against observed reality so as to inform a new, better prediction. Retesting, with iteration *ad infinitum*, inevitably leads to an improved understanding of evolutionary causes, both the specific selection pressures involved and the unexplained role attributable to random genetic drift (Parker & Maynard Smith, 1990). Traditionally, this approach has been used with organismal phenotypes, especially behavior, such as the time dung flies spend mating (Parker, 1970), or what size mussel a shore crab chooses to crack open for food (Elner, 1978). The optimality approach has been adapted, however, to molecular fundamentals: first to the size and content of the genetic alphabet (Szathmary, 1991; Szathmary, 2003), then the distribution of amino acid "meanings" within the genetic code itself (Freeland & Hurst, 1998; Omachi et al., 2023) and now to the amino acids as one possible set among many (Philip & Freeland, 2011; Mayer-Bacon & Freeland, 2021).

      A primary challenge for amino acids is to define and quantify features of their chemical structures upon which natural selection could plausibly have acted, not only for the canonical twenty but also for xeno alternatives. Careful biophysical measurements of amino acids from beyond the standard alphabet form an excellent example of where current science offers little data. It is not that it is difficult to identify biophysical aspects of protein folding consistent with what we observe in nature. However, correlation does not imply causation and there is currently little evidence with which to test this understanding other than a single, pioneering database of short, human-engineered peptides which each contain ~one xeno or non-coded amino acid (ModPep: see Singh et al., 2015). By the 21st century, however, computational chemistry was creating algorithms that could predict fundamental biophysical properties of molecular structures, and these were shown to be fully capable of estimating accurately the properties of molecules the size and complexity of L-$\alpha$ amino acids (Lu and Freeland, 2006). From here, investigations of the ways in which the genetically encoded amino acids distinguish themselves from xeno alternatives needed only the addition of one more, elegantly simple idea from the 21st century: chemistry space (Dobson 2004; Lipinksi & Hopkins, 2004). Chemistry space thinking is that any carefully defined, measurable biophysical property of a chemical structure may be thought of as a coordinate, such that measuring several properties of one molecule begins to define its chemical structure as a point within a multidimensional space. Equivalent measurements for other molecules define a cloud of points, wherein proximity means similarity, and distance means dissimilarity. Any statistical and/or geometrical concepts to compare points can be used to test quantitative hypotheses. The application of such thinking to organic chemical structures quickly revolutionized pharmacology, particularly the drug discovery industry (Lipinksi, 2016)

      By 2011 analysis of the chemistry space occupied by amino acids detected a highly unusual distribution by comparing the genetically encoded 20 with xeno alternatives.The two specific physicochemical properties involved (volume and hydrophobicity) are known to guide protein folding (Lins & Brasseur, 1995; Vascon et al., 2020). Subsequent work has expanded this evidence in both depth (Mayer-Bacon & Freeland, 2021; Mayer-Bacon et al., 2022) and breadth (Ilardo & Freeland 2014), and has even detected strong, non-random patterns in additional amino acid properties (Granold et al., 2018; Moosmann, 2021). The most recent expansion of this work has now identified for the first time specific examples of entirely xeno amino acids that match the statistical profile established by life since LUCA (Brown et al., 2023). An exciting next step will be to test empirically whether such alphabets exhibit some sort of identifiable advantage for protein folding.

# Discussion

This review synthesizes current knowledge regarding three, overlapping questions: (1) Would xeno biochemistry use amino acids? (2) Would it use monosubstituted L-$\alpha$-amino acids? and (3) Would it use different side chains? Below we summarize answers to each, along with examples of tractable near-future milestones of particular relevance to astrobiology.

**Would xeno biochemistry use amino acids?** One set of 20 amino acids has allowed life on Earth to inhabit an impressive diversity of environments. Indeed, conditions now recognized as supporting life on Earth overlap considerably with those identified for other planetary bodies in the solar system (Cockell et al., 2016; Merino et al., 2019). On another front, simulations and meteorite analyses agree that amino acids form readily under a wide range of abiotic and prebiotic conditions. Other organic polymers (e.g. hydroxy acids) that also form readily, polymerize with bonds that are less stable to hydrolysis than the peptide bonds which link amino acids. Depsipeptides thus spontaneously self-purify towards greater amino acid enrichment under wet-dry cycling. In other words, organic chemistry offers good reasons why any life might be expected to "encounter" amino acids (Fig. 8: outer shell), chemistry provides good reasoning why it might use them and biology shows us how resilient and versatile would be the result.

**Would a xeno biochemistry use monosubstituted L-α-amino acids?** Without catalysis, undirected chemical syntheses of amino acids generally produce equal amounts of L- and D- enantiomers. Even in the rare cases where L-enantiomeric excess has been detected, it is bias rather than absence of the D-enantiomer. At present, the most plausible inference is that genetically encoded homochirality arose through at least some evolutionary feedback, whether chemical or biological (see Wong et al., 2023 for an exploration of the difference). Such thinking introduces a new perspective beyond physicochemical arguments for using amino acids at all. Of course, even biological evolution could imply natural selection or genetic drift. So far evidence is stronger for natural selection and the current debate about L- chiral and α-amino acids is therefore better characterized as negotiating the relative power of biophysical constraint versus natural selection (Figure 8, middle shell).

Candidates to drive selection are not hard to identify, at least in some ways. Using only one stereoisomer permits the foundation of all protein structure as we know it: protein secondary structure is stabilized through intramolecular interactions that would be obstructed for a polymer comprising heterogeneous stereoisomers (Fig. 4). It is entirely reasonable, then, that biological evolution selected a homochiral set of amino acids for efficiency. However, multiple overlapping interpretations of efficiency are easy to think of and it remains to be determined which form the best guide to understanding how life on Earth has turned out. Similar thinking addresses the number of side chains per amino acid. Peptides comprising disubstituted amino acids can produce secondary structure, but they would inevitably require more energy for synthesis, transport, manipulation, and degradation due to their greater mass. This is the logic of the "Lego Principle": "*General arguments of thermodynamic efficiency … suggest that selectivity [to a reduced set of molecular building blocks] is required for biological function and is a general result of natural selection*" (McKay, 2004).

From here, current understanding of a third feature for amino acids proceeds easily. The genetically encoded set of amino acids are all ***α***-amino acids. Again, prebiotic simulations and meteorite analyses clearly indicate that amino acids with longer backbones (β-, γ-, and δ- amino acids etc.; Fig. 3) were available throughout life's origin, and the widespread use of these longer-backbone amino acids within current biology demonstrate their continued availability throughout evolution. It has long been recognized that each carbon-carbon bond within these longer amino acid backbones presents a new site of possible rotation and that the resulting increase in flexibility for peptides, reducing structural stability. Fifty years ago, it seemed clear that this biophysical constraint accounted for evolution's "choice" of alpha amino acids. Once again, subsequent research has demonstrated that longer backbones and/or even a more heterogeneous diversity of backbone lengths can produce viable and even biomedically relevant protein structures (reviewed in Cabrele et al., 2014). It is at least as compelling then to suggest that evolution, and specifically natural selection, would have favored the lower energy budget of working with the smallest, least massive backbone as a universal feature of its monomeric building blocks.

**Would a xeno biochemistry use different side chains?** As was true for alternative backbones, we now know that many other side chains can function within the genetic code, even after 3.5 billion years of evolution (Hickey et al., 2023). *De Novo* Protein Design is teaching us how to create and analyze peptides and proteins that incorporate xeno amino acids, providing tools and techniques with which to move into this uncharted territory (Makarov et al., 2023). The main contribution of prebiotic chemistry, in the process of connecting the abiotic universe to life, has been providing empirical evidence that alpha amino acid backbones cost less to make and use (Cleaves, 2010). Molecular Evolutionary Biology has started to develop specific predictions for alternative amino acid alphabets by identifying and emulating quantifiable, biophysical properties of the encoded 20 (Mayer-Bacon et al., 2021).

**In summary:** Current knowledge about alternatives to the L-$\alpha$-amino acids used by life indicates that as the focus narrows from amino acids as a chemical class to the specific 20 side chains genetically encoded by post-LUCA biology, the influence of biophysics diminishes relative to that of biological evolution (Fig. 8). That being said, much remains unknown. Whereas Weber and Miller (1981) concluded that "*we would expect that the catalysts would be poly-alpha-amino acids and that about 75% of the amino acids would be the same as on the earth*", in 2023 we suggest instead that emerging ideas, technologies and datasets are positioned to make such estimate possible within the next decade.

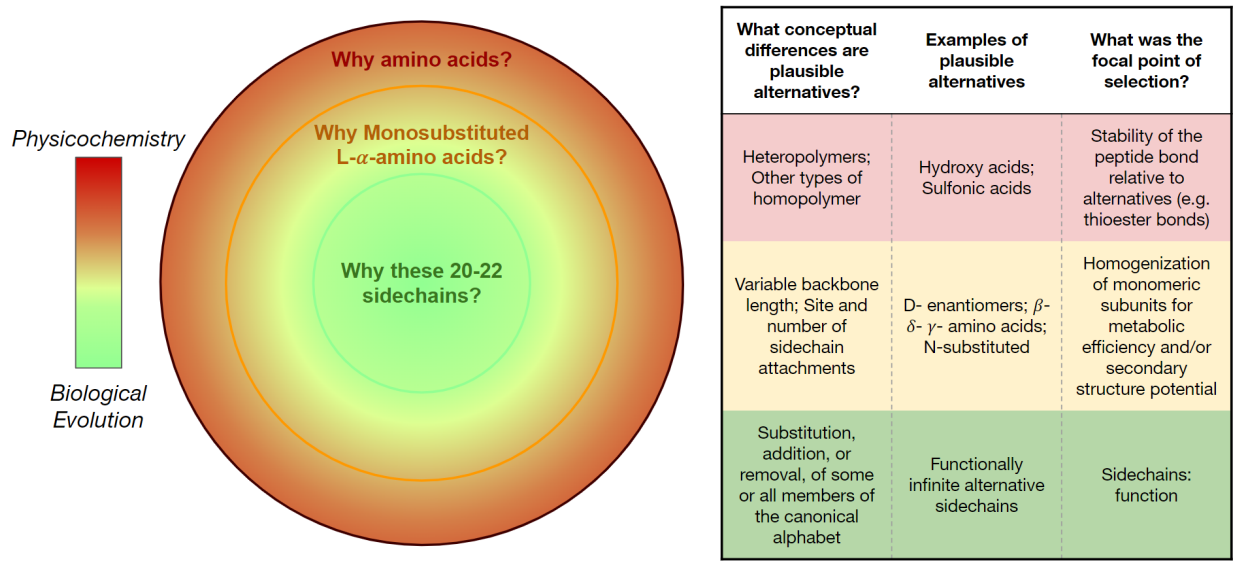

**Figure 8:** Why does life on Earth use one precise set of 20 L-$\alpha$-monosubstituted-amino acids? As the focus of this question narrows from amino acids as a class of chemicals to the 20 specific side chains used by post-LUCA life, the probable role of physicochemical constraint diminishes relative to that of biological evolution.

**What tractable questions would represent progress for xeno amino acid science?** From this conclusion, we identify 3 overlapping near-term goals that will expand our understanding of biochemistry as we don't know it.

Regarding alternatives to amino acids as monomeric building blocks, the current frontier is characterizing their potential to polymerize. Looking ahead, it is rapidly becoming both tractable and important to expand current work (e.g. Chen et al., 2022) that characterizes structures and functions of individual examples into general, systematic statements about how such xeno polymers differ from proteins.

Similarly, the design and synthesis of amino acid polymers with longer backbones ($\beta$-, $\gamma$-, $\delta$-, etc.), mixed chirality, or multiple side chains are increasingly well understood to form structures and functions (e.g. Forsythe et al., 2018). From here, it will be exciting to see systematic characterization that quantifies the structural and functional range of such molecules relative to L-$\alpha$-amino acid polymers.

Finally, for side chains momentum is growing for biomedical research that engineers noncanonical amino acids into otherwise natural proteins (Hickey et al., 2023), and for work that mixes canonical and xeno side chains from an origins perspective (e.g. Makarov et al., 2023; Fenkel-Pinter et al., 2019). From here a natural new milestone will be to see these approaches meld into rational design of entirely xeno proteins and even xeno alphabets. For this to happen, one of several innovations needed is for two branches of theory to connect: sophisticated biophysical calculation of structure applied to xeno alphabets designed to emulate biology's canonical alphabet. Both approaches seem likely to learn and grow from each other.

## Author Contributions

Conceptualization, Resources, Data Curation, Writing — original draft preparation: Sean Brown and Stephen Freeland. Writing — review and editing: Sean Brown, Stephen Freeland, and Christopher Mayer-Bacon. Funding acquisition: Stephen Freeland. All authors have read and agreed to the published version of the manuscript.

## Acknowledgments

We would like to thank Ashley Copenhaver for constructive dialog about amino acids' roles within neurotransmission, Jessie Novak and Julia Sunnarborg for information about D-amino acids' role in neurotransmission and venom, Robin Kryštůfek for equally helpful dialog about empirical protein biochemistry, Bonnie Teece for similar expert consulting regarding meteoritic abundances of organics, Erin Gibbons for minor edits regarding exoplanet and solar system environments, Valerie Zhou for improving our figures, and Corleigh Forrester for help with figure legends.

Marvin was used for drawing, displaying, and characterizing chemical structures, substructures, and reactions, Marvin 22.22, 2023, ChemAxon (http://www.chemaxon.com)

## Funding

This work is funded through the Human Frontiers Science Program (HSFP) RGEC27/2023 Research Grant and the University of Maryland, Baltimore County: Department of Biological Sciences.

## Conflicts of Interest

Christopher Mayer-Bacon contributed to this article in his personal capacity. The views expressed are his own and do not necessarily represent the views of the FDA or the United States government.